# An Effective Black Hole Attack Detection Mechanism using Permutation Based Acknowledgement in MANET

Dhaval Dave
CSE Dept.
NIT - Warangal
davedhaval87@yahoo.com

Pranav Dave
CSE Dept.
LDRP - Gandhinagar
pranavdave893@gmail.com

*Abstract*- **With the evolution of wireless technology and use of mobile devices, Mobile Ad-hoc Network has become popular among researchers to explore. A mobile ad-hoc network (MANET) is a self-configuring network of mobile routers (and associated hosts) connected by wireless links. The routers and hosts are free to move randomly and organize themselves arbitrarily. It allows mobile nodes to communicate directly without any centralized coordinator. Such network scenarios cannot rely on centralized and organized connectivity, and can be conceived as applications of Mobile Ad-Hoc Networks. Thus, MANET is vulnerable due to its dynamic network topology, as any node become untrusted at any time. The Black hole attack is one kind of security risk in which malicious node advertises itself to have a shortest path for any destination, to forge data or for DOS attack. In this paper, to detect such nodes effectively, we propose a Permutation based Acknowledgement for most widely used reactive protocol ad-hoc on demand distance vector routing AODV. This mechanism is enhancement of Adaptive Acknowledgement (AACK) and TWO-ACK, here we have tried to show the efficiency increment by decreasing number of messages routed in the network.**

## I. INTRODUCTION

Mobile Ad-hoc Network is collection of anonymous nodes communicates with each other in decentralized manner without the need of any centralized router or server and without help of any fixed topology. All nodes are self-configure and communicate over relatively bandwidth constrained wireless links. Since the nodes are mobile, the network topology may change rapidly and unpredictably over time. As network is decentralized, all network activity including discovering the topology and delivering messages must be executed by the nodes it selves, i.e., routing functionality will be incorporated into nodes.

As MANET starts gaining popularity, the security issues [1] has become one of the primary concerns because MANET is vulnerable to various type of attack such as passive or active attack. Passive attack does not affect the communication over the network it only read the confidential information. Active attack affects the normal operation of network by altering it. A Black hole attack is an Active attack, which intentionally drops the packet. It is a malicious node that attracts all packets by using forged RREP to falsely claiming a fresh and shortest route to the destination and then discards them without forwarding them to the destination. This is depicted in Fig. 1. The main purpose of malicious node is to disturb the communication over the network and affect data availability.

In this paper, we propose Ad-hoc On-demand Multipath Secure Routing (AOMSR) a Black hole node detection system for mobile Ad-hoc network using a Permutation Based Acknowledgment (PBAck). This technique uses Ad-hoc Distant Vector Routing (AODV) protocol to achieve this goal. AODV is used because it is a simple and efficient routing protocol designed specifically for use in multi-hop wireless ad-hoc network. This mechanism solves the Black hole node problem using a less number of broadcast messages, as compared to other proposed and used technique so far.

## II. RELATED WORK

In MANET, Security issues have always been a hot topic and many research monograms / articles are available in the literature that deals exclusively the black hole problem. Following are related works that are done previously:-

Lu et al. [2] proposed SAODV which may be considered as an extension to the AODV. On receiving a RREP packet, the source node verifies a secure path to the destination node by sending SRREQ packets. The SRREQ contains a secret code (a random number). After receiving at least two such packets, the destination node responds back with SRREP packets. The SRREP also contains a secret code (a random number). When the source node receives at least two such packets, it chooses the shortest among them as a secure path to the destination node for data transmission.

Deng, Li and Agrawal et al. [3] also propose a method which verifies security of the path after receiving a RREP packet. It requires each node to send back the next hop information also when it sends back the RREP packet. After receiving the RREP packet, the source node sends a FurtherRequestpacket to the next hop of the intermediate node. Only the next hop can send back a FurtherReply packet. The source node decides a secure route on the basis of this FurtherRequest packet and declares the malicious node as black hole. The drawback of this approach is an increased delay in the network.

Alem and Xuan [4] propose a solution Intrusion Detection using Anomaly Detection (IDAD). It uses host-based Intrusion Detection System (IDS) scheme to monitor the activities of a host. An anomaly activity is detected on the basis of audit data which is collected and is given to the IDAD system. It compares every activity of a host with the audit data on the fly and isolates a host (node) if any of its activity resembles an activity in the audit data. However, there are several drawback of this method. It requires extra memory, slows down the system and is impractical to implement in some hostile scenario.

Venkatraman and Agrawal [5] present an external attack prevention and internal attack detection model for AODV. The external attack prevention model secures the network from external attacks by implementing message authentication code to ensure integrity of the route request packets. The internal attacks are detected by internal attack detection model which uses traces of local data to identify and isolate the misbehaving nodes. However, it has a high false positive rate as sometimes it is difficult to differentiate abnormal behavior with normal behavior.

Pushpa [6] presents a modified approach of AODV which is based on trust and gives equal weight to both route trust and node trust for the route selection process. Continuous evaluation of node's performance and collection of neighbor node's opinion value about the node are used to calculate the trust relationship of this node with other nodes.

Medadian et al. [7] present a routing protocol to combat black hole attack in MANET. It is a trust based method where the sender takes opinion of the neighbors of the node (say, node A) which replied with a RREP packet, i.e., advertises the shortest route to the destination. This opinion along with a rule base determines whether node A is malicious.

Mahmood and Khan [8] present a survey of methods to combat black hole attack on AODV routing protocol.

Pramod and Govind [9] propose a comparatively simple and efficient method to detect and isolate black hole in a network with minimal routing overhead.

Botkar and S.R. Chaudhary [10] have proposed AACK, which is Advanced Acknowledgment, it may be consider as combination system of an Enhanced TWOACK (E-TWOACK) scheme and an End-to-End Acknowledgment scheme. The difference between the E-TWOACK and TWOACK [14] is that in TWOACK technique it detects the malicious link and, in E-TWOACK technique it detects the malicious node instead of link that's why the detection efficiency gets increased.

## III. BASIC CONCEPTS:

A. AODV Routing Protocol: Ad-hoc on demand distance Vector Routing (AODV) [11] is a simple and efficient routing protocol designed especially for use in multi hop wireless Ad-hoc network. AODV is a reactive protocol in which routes are created only when they are needed. It uses traditional routing tables, one entry per destination, and sequence numbers to determine whether routing information is up-to-date and to prevent "counting to infinity" problem.

When a node wishes to transmit data to a host to which there is no route, route discovery will be carried out, it will generate and broadcast routing request message (RREQ), routing reply message (RREP) is unicasted back to the source of RREQ, and route error message (RERR) is sent to notify other nodes of the loss of the link. HELLO messages are used for detecting and monitoring links to neighbors.

B. Black Hole Attack: A black hole attack [12] is a type of denial of service attack that can be easily employed against routing in Ad-hoc networks. In this attack, a malicious node tries to attract all packets by advertises itself as having the shortest path to the any destination node.

When the attacker node receives, an RREQ message, without checking its routing table, immediately sends a false RREP message giving a route to destination over itself after hop count value is set to lowest values and the sequence number is set to the highest value to settle in the routing table of the victim source node. Therefore source node assumes that route discovery process is completed and ignores other RREP messages and begins to send packets over attacker node. Attacker node attacks all RREQ messages this way and takes over all routes. Therefore all packets are sent to a path, they are simply dropped and will not be reached to appropriate destination.

## IV. PROPOSED CONCEPT

We propose a Ad-hoc On-demand Multipath Secure Routing (AOMSR) method which uses Permutation Based Acknowledgement (PBAck) to detect the malicious node. A Source needs to store multiple paths from source to destination based upon maximum delay endured in receiving data.

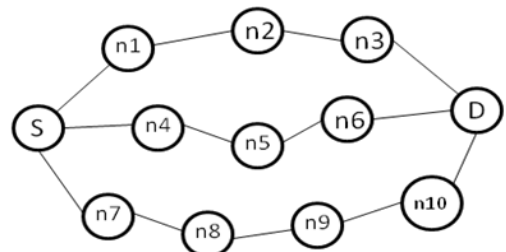

Figure1: Assume MANET system

Detailed Process is as follows. Consider a Source node S and Destination node D. There are several paths from S to D. but based upon shortest path estimated time "t", and maximum allowed delay from "t+δ", S chooses K paths from S to D, here in example we choose K=3.

| Path Number | Permutated Acknowledgement Number | Number of Path | Type Of Message |
|---|---|---|---|
| PN | PAckN | NP | ToM |

Figure 2: Appended data structure to Message header

| Path Number | Permutated ACK Number | Path | Path Good / Bad | Path Free | PBA sent / received |
|---|---|---|---|---|---|
| PN | PAckN | P | FGB | FFB | PSR |

Figure 3: Data Hash Table (DHT)

Figure 2 is showing a data structure appended to the message header, which will indicate the message coming from the which path (PN), this message's permutated acknowledgement (PBA) should be sent from which path (PAckN), total number of path which can be useful to receiver to understand making Data Hash Table (DHT), and Type of Message ToM, which is significance of the message type either Data or Path.

Figure 3 is representing a data structure called Data Hash Table (DHT), which will be created at source and destination node, to keep track of data packets received and its PBA is sent or not.

Sender S will send different data packets via different paths with path number (PN) and Permutated Acknowledgement Number (PAckN) such that PN and PAckN are not same number to destination, at the same time S will keep all sent data packet's PN, PAckN, Path, and several flags like flag for path is good or bad (FGB), Flag for path is Free or Busy (FFB) in a table. In this example three data packets will go via three paths with PN=1, 2, 3 & PAckN=3, 1, 2 respectively.

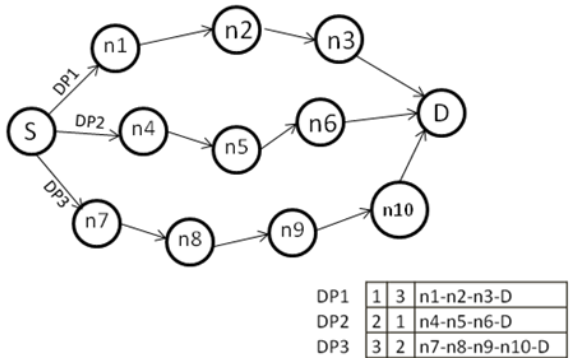

Figure 4: Data packet flow

Destination D receives data packet stores its all relevant entries in a table. And sends Permutated Acknowledgement to destination via pre decided paths.

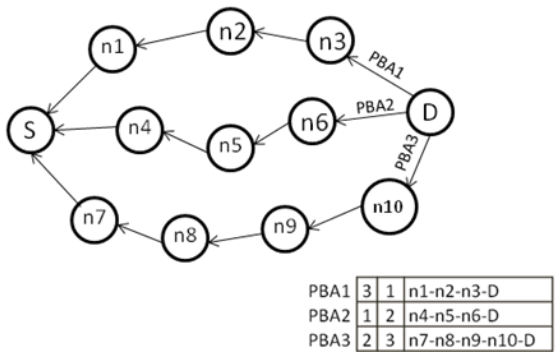

Figure 5: PBA packet flow

In this example data packet 1's Acknowledgement has to come from path on which data packet 2 is sent, because data packet 2 has PAckN as 1, same way data packet 2's PBA has to come from path 3 and data packet 3's PBA has to come from path 2, if destination receives all data packets correctly, then it will send all Permutated Acknowledgement PBA correctly.

Consider a scenario where a node is malicious node and drops the packet, than Destination D will not have its PN, PAckN, and Path so D will not be able to send its Permutated Acknowledgement PBA to different path and some received data's PBA to that path. So Destination will not receive two PBA for one data lost, for example if node "n6" in path 2 is Black hole attacker it has no connectivity to Destination and drops the packets, data packet with PN as 1 and PAckN as 3 reaches destination correctly, data packet with PN, PAckN as 2,3 does not reaches to Destination and data packet with PN, PAckN as 3,2 is received , as soon as Destination receives data packet form path 3, it will send on this path PBA of data packet 1, and waits for data packet with PN, PAckN as 2,1. At sender side till timer goes off, only one PBA of data packet 1 have been received via path 3 so sender will understand that data packet 1 and 3 have been received correctly as PBA of 1 is received and form path 3, and as PBA of 3 via path 2 is not received and PBA of 2 via path 1 is not received so there can be Black hole attacker node in path 2.

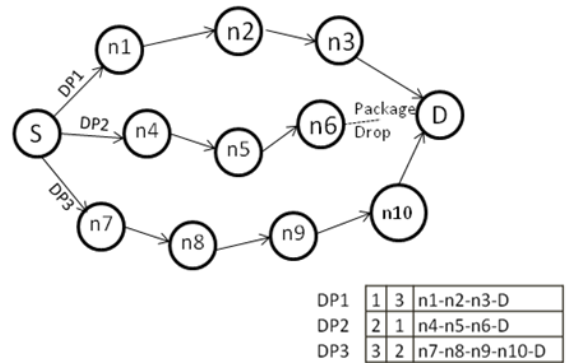

Figure 6: System with Black hole node

Now Sender will send path which may contain Black hole attacker node "S-n4-n5-n6-D" in a data packet setting Type of Message (ToM) as 0 via path 3, because it is marked now as Good path. Destination will receive data in which path is there, and check for its node which has announced to have connectivity with Destination D, here n6, if D has no neighbor as n6, then it will be marked as Black hole attacker node and Alarm packet will be broadcasted.

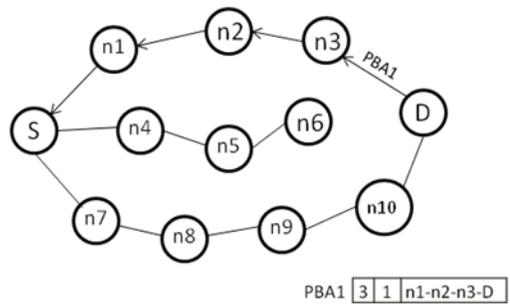

Figure 7: PBA in system with Black hole node

## V. ASSUMPTION

While proposing this concept we have assumed some factors which are as below:

A black hole attacker will announce its direct connectivity with Destination to reduce the hop count and maximize the sequence number.

There is no cooperative black hole attack such that all attackers from different paths, chosen by sender can communicate with each other without sending data to destination otherwise they compromise and send a forged PBA to source node, and source node will never know about existence of black hole attacker and Destination will never get data packets

## VI. ALGORITHM

| | |
|---|---|
| Source node: | S |
| Destination node: | D |
| Route Request: | RREQ |
| Route Reply: | RREP |
| Modified Routing Table: | MRT |
| Destination Sequence Number: | DSN |
| Hope Count: | HC |
| Flag path good/bad: Good-1 Bad-0 | FGB |
| Flag path free/busy: Free-1 Busy-0 | FFB: |
| Path | P |
| Path Number: | PN |
| Permuted Acknowledge Number: | PAckN |
| Permuted Acknowledge: | PBA |
| Total number of Paths: | NP |
| Type of Message: Data -1, Path-0. | ToM |
| Modified Routing Table | MRT |
| Data Hash Table: | DHT |
| PBA sent / received | PSR |
| Intermediate nodes: | n1, n2 … nk. |
| Assumed Source to Destination Data delivery time | T |
| Assumed estimated Source to Destination Data delivery time via shortest path: | Tmin |
| Maximum delay acceptable than estimated time: | δ |

Table 1: Parameters used in the algorithm

*Sender to Receiver:*

**STEP 1:** GeneratePath() : Desired number of path creation
**Require:** Auxiliary ''S'' node wants to start transmission of data to destination ''D'' node, for which it don't have path.

1. S broadcasts RREQ packet with destination address of D in the network.
2. *if* RREP received
2.1. Set in MRT ← DSN, HC, P
3. *else* go to 1.
4. *end if*

**STEP2:** SelectMultiPath()
**Require**: S has to select paths to send data from all paths stored in MRT
1. PN←0, NP←0;
2. *while (1)*
2.1. KeepP whose T ≤ $T_{min}$ + δ.
2.2. Assign PN to P.
2.3. NP = NP+1.
2.4. PN=PN+1.
3. *end while*

**STEP 3**: GeneratePermuatatedPath(NP)
1. *for* 1 to NP
1.1 generate a permutated sequence of PAckN such that PN ≠ PAckN.
2. *end for*

**STEP 4:** FormDataPacket(PN,PAckN,NP,P)
**Require:** S produces data packets DP.
1. Set PN, PAckN, NP, ToM=1 to message header
2. Set in DHT ←PN, PAckN, P, FFB=0, FGB=0.
3. Send DP to path PN.

**STEP 5:** ReceivesPBAck(PBA):
**Require**: When S receives Acknowledgement PBA
1. search entry E such that PBA's PN =E's PN in DHT.
2. *if* E is found
2.1. set E's PSR ←1, FGB ← 1, FFB ← 0.
2.2. search entry E2 such that PBA's PAckN = PN.
2.3. set E2's FGB ←1 , FFB ← 0.
3. *end if*
4. *else* discard PBA.

**STEP 6:** TimerGoesOff ( ):
**Require**: Sender's timer for acknowledgement goes off
1. *if* timer goes off
1.1. search entry E in DHT such that PSR = 0 && FGB = 0.
1.2. *if* E is found.
1.2.1. form a DP with data as path of E.
1.2.2. ToM ←0.
1.2.3. Send DP via smallest path which has FGB = 1.
1.3 *end if*
1.4 *end if*

*Receiver to Sender:*

**STEP1**: PocessDataPacket ( DP )
**Require**: D receives data packet DP, stores its PN, PAckN, and Path P in DHT,
1. *if* ToM ≠ 0
1.1 search for entry E in DHT such that DP's PN = PAckN in DHT.
1.2 *if* entry E is found .
1.2.1 Send PBA to path of E.
1.2.2 Set PSR ←1.
1.3 *end if*
2 search for entry E in DHT such that DP's PAckN= PN in DHT.
3 *if* entry E is found && E's PSR = 0.
3.1 send PBA of E's Data to Path of received DP.
3.2 set PSR← 1.
3.3 set FGB ← 1, FFB ←1.
4 *end if*
5 *else if* ToM = 0
5.1 Destination stores data as path and checks the node which has announced direct connectivity to D

5.2 *if* node does not exist in D's neighbor list than
5.2.1 Mark node as Black hole attacker node,
5.2.2 broadcast Alarm Packet.
5.3 search for this path P in DHT.
5.4 Set FGB as 1.
5.5 Reply this to Sender S.
5.6 *end if*
6. *end if*

## VII. ANALYTICAL RESULTS

To find the efficiency of our proposed concept to find the Black hole attacker in a MANET system, we compare with the concept proposed in [10]

In paper [10], if "n" is the number of nodes in a MANET system within the presence of Black hole attacker node.

In absence of black hole attacker node in network, for each data sent form sender S to destination D, One "End to End Acknowledgement EEACK" will be sent from D to S, which will cause "n" transaction of intermediate nodes of EEACK.

If EEACK has not been received due to black hole attacker node or link failure or damaged ACK packet. System will start running in to E-TWOACK [13] mode which overall performs "2*n" transactions in between nodes as in following example 4 nodes will create 8 ACK transactions.

In below example there are four intermediate nodes, which will cause eight ACK transactions.

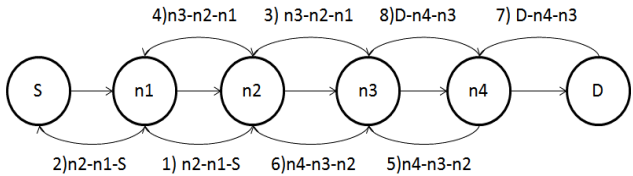

Figure 8: E-Two-Ack working and transaction in four nodes

So on an average there will be 3n transactions (2n for E-TWOACK +n for EEACK) for detection of black hole attacker node.

Total number of ACK transfers = 3n

In our proposed system
Let's assume:
Number of Data path = k
Maximum Intermediate nodes form S to D in each path= n.
Total PBA transactions =k*n.
where, Per data "n" PBA will be sent by destination.

If any PBA has not been received and its path is flagged as Bad, corresponding path will be sent to Destination D via any Good flagged path (FGB=1), which will cause another "n" data transactions.

So, total "2n" communications will be carried out find Black hole attacker node efficiently. Fairly which is better than [2] by 33.33% because instead of using bandwidth by 3n we use 2n transactions and Detecting malicious node is more accurately.

RO = Routing_Overhead
CP = Control_Packet
DP = Routing_Packet

$$RO = \frac{CP}{CP + DP} * 100\%$$

With above calculations as proposed in [2], CP = "2n" (of E-TWOACK), DP = "n" so RP = 66.66%.
In proposed scheme

For "p" malicious path out of "k" available paths p < k.
CP = "p*n" , DP = "k*n" , where k is number of data path .

$$RO = \frac{pn}{pn + kn} = \frac{p}{p + k} * 100\%$$

For different values of p and k calculation table is as follows:

|  | K = |  |  |  |  |
|---|---|---|---|---|---|
| P = | 1 | 2 | 3 | 4 | 5 |
| 1 | 50.00 | 33.33 | 25 | 20 | 16.66 |
| 2 | -NA- | 50 | 40 | 33.33 | 28.57 |
| 3 | -NA- | -NA- | 50 | 42.55 | 37.5 |
| AACK | 66.66 | 66.66 | 66.66 | 66.66 | 66.66 |

|  | K = |  |  |  |  |
|---|---|---|---|---|---|
| P = | 6 | 7 | 8 | 9 | 10 |
| 1 | 14.28 | 12.5 | 11.11 | 10 | 9.09 |
| 2 | 25 | 22.22 | 20 | 18.18 | 16.66 |
| 3 | 33.33 | 30 | 27.27 | 25 | 23.07 |
| AACK | 66.66 | 66.66 | 66.66 | 66.66 | 66.66 |

Table 2: Analytical values of Routing Overhead for different malicious paths & total paths

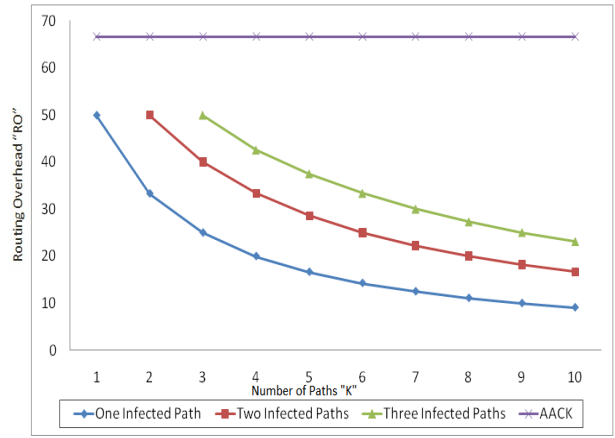

Figure 9: Routing overhead vs. Number of path

Figure 9 shows theoretical comparison of AOMSR and AACK for routing overhead. AOMSR uses multiple paths and hence routing overhead decreases as number of paths increase in steady simulation without path breaks.

## VIII. SIMULATION ENVIRONMENT

Proposed method has been implemented in NS2. The simulation consists of different number of nodes placed randomly in rectangular region. For data transmission Constant Bit Rate (CBR) transmits UDP based traffic over random mobility model of nodes with 2MBPS bandwidth, each data packet is 512 byte long. Mobile nodes are assumed to move randomly as per random way point model [15].

## IX. RESULTS

To measure two performance metrics of *Throughput* and *Routing Overhead* two simulation environments are chosen with constant mobility and variable mobility over traditional AODV and AOMSR implementation.

Network throughput = Number of data packets Received/ no of data packets Sent.

Routing Overhead = Total number of control packets transmitted by nodes while establishing and maintaining routes.

### A. CONSTANT MOBILITY

Mobility is constant between 0 and maximum speed 10 m/s with pause time 30 seconds for each simulation. Simulation time is 300 seconds.

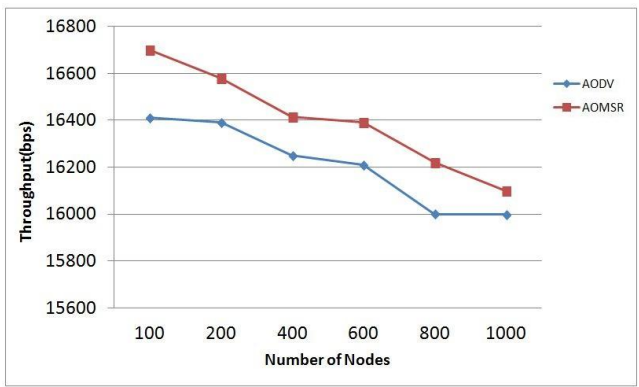

Figure 10: Throughput (bps) vs. Number of Nodes

Figure 10 shows comparison between AOMSR and AODV with respect to throughput. Number of packets received is decreasing as network size and number of network increases, due to increasing number of path breaks. Our proposed AOMSR increases the packet delivery due to availability of multiple paths.

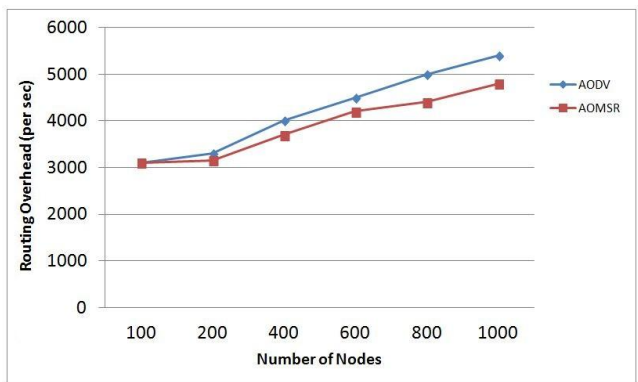

Figure 11: Routing Overhead vs. Number of Nodes

Figure 11 represents Routing Overhead comparison between AOMSR and AODV with respect to number of nodes. Theoretically AOMSR outperforms AACK, practically we compared proposed scheme with AODV. AOMSR gives lower overhead in larger network due to availability of multiple paths while AODV performs route computation when route break detected.

### B. VARIABLE MOBILITY

Number of nodes is between 10 to 50 and mobility kept varying between 10 to 50 m/s. Simulation time is 300 seconds. Average numbers of all observations are taken to generate graphs.

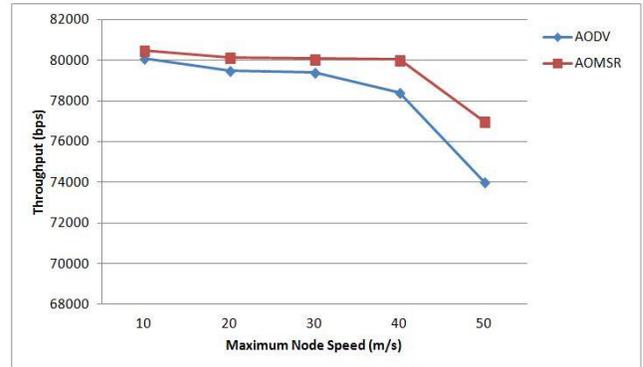

Figure 11: Throughput (bps) vs. Node Speed

Figure 11 represents throughput with respect to node speed. Probability of link failure increases with high node speed, hence as shown, with high node speed throughput decreases in AOMSR as well as AODV.

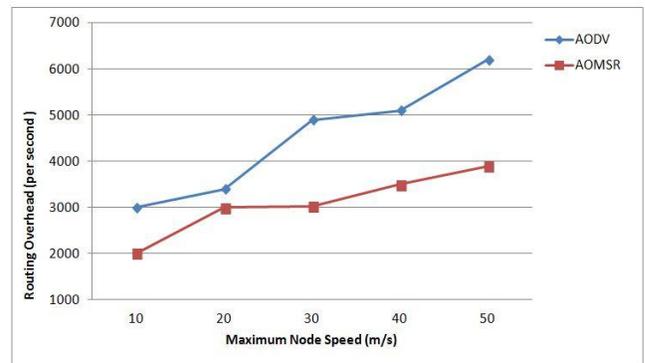

Figure 12: Routing Overhead vs. Node Speed

Figure 12 represents relationship between routing overhead and Node speed. With high node speed routing overhead increases in AOMSR and AODV significantly.

## X. FUTURE WORK

The simulation of this project can also be conducted by varying topologies on the basis of factors: Behavioral properties of node (the number of black hole attacker and good nodes involved in communication). Other factors such as mobility speed of the nodes, load over network, priority of data packets etc. may also be varied to evaluate their impact on a protocol's performance. The experimental scenarios and findings can also be verified by using other simulation tools like GloMoSim, NCTUns, Omnet++ etc.

Throughout this project, we have chosen Throughput, number of packets as overhead. Other parameters such as end-to-end all packets' delay, efficiency of finding black hole attack or packet delivery fraction may be used to compare performance.

The detection of malicious node mechanism for AODV has the most potential for improvement and significant work has been done and proposed in this area. We have chosen the multiple paths routing to detect black hole attacker efficiently, to reduce end-to-end delay and to increase throughput.

## XI. CONCLUSION

MANET is an increasingly emergent field in which various researches are putting their effort to solve the problem of various attacks in it like Black hole attack. MANET has great potential in various diverse areas, e.g., military, disaster management, intelligent transportation system, monitoring, public safety. However, it poses a greater security risk in comparison to conventional wired and wireless networks due to its inherent properties. In this paper, we propose a simple, efficient and effective method with minimum routing overhead and good throughput to combat the black hole problem. It does not require any database, extra memory and more processing power. The simulation and results clearly mention about accuracy of AOMSR even theoretically & analytically our propsed scheme is better and more effective than others. In Future, we put our best efforts to make it more efficient to find Black hole node.